# Regret Minimizing Equilibria and Mechanisms for Games with Strict Type Uncertainty

**Nathanael Hyafil**
Department of Computer Science
University of Toronto
Toronto, ON, M5S 3H5, CANADA
nhyafil@cs.toronto.edu

**Craig Boutilier**
Department of Computer Science
University of Toronto
Toronto, ON, M5S 3H5, CANADA
cebly@cs.toronto.edu

## Abstract

Mechanism design has found considerable application to the construction of agent-interaction protocols. In the standard setting, the type (e.g., utility function) of an agent is not known by other agents, nor is it known by the mechanism designer. When this uncertainty is quantified probabilistically, a mechanism induces a game of incomplete information among the agents. However, in many settings, uncertainty over utility functions cannot easily be quantified. We consider the problem of incomplete information games in which type uncertainty is *strict* or unquantified. We propose the use of *minimax regret* as a decision criterion in such games, a robust approach for dealing with type uncertainty. We define *minimax-regret equilibria* and prove that these exist in mixed strategies for finite games. We also consider the problem of mechanism design in this framework by adopting minimax regret as an optimization criterion for the designer itself, and study automated optimization of such mechanisms.

## 1 Introduction

As software agents become better able to act on behalf of human users, organizations, and businesses, there is an increasing need to develop environments in which such agents can interact smoothly, and protocols to ensure that desired outcomes can be reached. For example, the migration of many day-to-day business transactions to on-line market, bargaining, and negotiation systems has led to the development of more and more sophisticated software agents that mediate such transactions. However, since the interests of the parties on whose behalf they act generally conflict, these agents should reason strategically according to the well-studied principles of game theory. Consequently, recent research in computer science and economics has focused on the design of economic agents and the mechanisms through which they interact.

*Mechanism design* [13] has played a central role in much of this research—it can be seen to embody the algorithmic principles of computer science within an economic context. Key results in mechanism design, such as the revelation principle, have had a strong influence on the direction taken by research at the intersection of the two disciplines. However, recently, limitations of standard approaches to mechanism design have been identified, and are starting to be addressed. Chief among these is the complexity of computation, communication, and the "human factors" faced by software agents. For instance, mechanisms based on the revelation principle require agents to reveal their *type* (generally, their utility function) accurately. This presents a problem in any of a number of different circumstances: (a) utility functions are often defined over large, multi-attribute spaces, and are difficult to communicate effectively and/or hard to compute accurately; (b) even in compact domains, obtaining precise utilities (e.g., the precise valuation for some good) may be computationally difficult (e.g., if the value of the good must be determined by solving a difficult optimization problem); and (c) the software agent may need to engage the user on whose behalf it is acting in order to obtain this utility, but this user may be uncomfortable specifying utility values to the required degree of precision.

Recent research has begun to examine methods involving limited or incremental elicitation of types to circumvent some of these difficulties [1, 7, 17, 18], specifically in the context of (single-good or combinatorial) auctions. But the question of mechanism design in general settings when only partial type information can be practically revealed has received scant attention. Furthermore, work to date (generally on auctions) has focused on "classical" mechanism design, in which type uncertainty is quantified probabilistically (with the designer and the participating agents having a common prior over types).

In this paper, we relax the form of the prior over types. Rather than quantified uncertainty, we assume that uncertainty over types is *strict*: all that is known about an agent is that its type lies within some set of possible types. This form of uncertainty might arise when, say, we know that the value an agent places on some good lies within certain bounds, but have no further probabilistic information about the valuation. In the context of mechanism design, an incremental or partial elicitation mechanism may further refine these bounds by having the agent reveal additional constraints without fully identifying its type.

To deal with this form of uncertainty, new solution concepts are required. In the standard formulation, a specific mechanism induces a *Bayesian (incomplete information) game* [13]. However, with strict type uncertainty, the nature of the induced game is different. In this paper, we adopt the notion of an *incomplete information game with strict type uncertainty* (these are equivalent to games in *informational form* [11]). Since the expected utility of a strategy profile in a strict incomplete information game cannot be defined, we propose a more qualitative decision criterion, *minimax regret*. By analogy with Bayes-Nash equilibria, we define a new solution concept—*minimax-regret equilibria*. Intuitively, a minimax-regret equilibrium is a strategy profile in which each agent minimizes its regret with respect to the realization of all other agent's types. We show that such equilibria must exist in finite games.

With minimax-regret equilibria in place, we then propose new criteria for mechanism design under strict type uncertainty. We argue that the *mechanism itself* should minimize its regret (in the social choice function), again with respect to realization of the types of participants. While this notion is applicable to direct mechanisms, we expect it will be especially critical for partial or incremental revelation mechanisms. We define novel mechanism design settings, and suggest various techniques by which the notion of *automated mechanism design (AMD)* [8, 20] can be applied to mechanisms with strict type uncertainty.

We begin with an overview of relevant background in Section 2. We then introduce strict incomplete information games in Section 3, and define minimax-regret equilibria. In Section 4 we address the issue of mechanism design under *partial type revelation* with minimax-regret equilibria, the form of optimization required for AMD, and report on preliminary experiments for the special case of complete type revelation. We conclude with a discussion of future research directions.

## 2 Background

In this section we provide some brief background on the key concepts used later in the paper, specifically, games of incomplete information, mechanism design, and minimax regret for decision making under strict uncertainty.

### 2.1 Games of Incomplete Information

Assume a collection of players $i \leq N$, with a set of actions $A_i$ for each player $i$. Let $A = \times_i A_i$ denote the set of *action profiles*. We assume a set $\Theta_i$ of possible *types* for each $i$, and a *utility function* $u_i : A \times \Theta_i \mapsto \mathbf{R}$. We assume throughout that $A_i$ and $\Theta_i$ are finite (though we discuss generalizations in several places below). Intuitively, the type of an agent captures all relevant aspects of the interaction, or game, that are private to that agent. Formally, an agent's type determines its utility for different action profiles (or game outcomes), since most forms of private information can modeled in this way [13]. As such we often speak as if an agent's type *is* its utility function. Let $\Theta = \times_i \Theta_i$ be the collection of *type profiles*. While each agent knows its own type, it is uncertain about the types of the others. This is reflected in a *common prior* over $\Theta$ where $\Pr(\theta)$ is the commonly known distribution from which type profiles are drawn. Agent $i$'s beliefs about the type profile $\theta_{-i}$ of other players, given its own type $\theta_i$, are given by the conditional distribution $\Pr(\theta_{-i}|\theta_i)$.

A *(Bayesian) game of incomplete information* is made up of the components above: players, actions, types, utility functions, and common prior. A *mixed strategy* for player $i$ is a mapping $\sigma_i : \Theta_i \mapsto \Delta(A_i)$, associating with each type $\theta_i$ a distribution $\sigma_i(\theta_i)$ over possible action choices. A strategy is *pure* if $\sigma_i(\theta_i)$ assigns probability 1 to a single action for each $\theta_i$; otherwise it is *strictly mixed*. Let $\Sigma_i$ denote the set of (pure or mixed) strategies for agent $i$, $\Sigma = \times_i \Sigma_i$ be the set of strategy profiles, and $\Sigma_{-i} = \times_{j \neq i} \Sigma_j$ be the set of profiles ranging over all agents except $i$.

Fixing the strategies of all other agents to $\sigma_{-i} \in \Sigma_{-i}$, the expected utility of strategy $\sigma_i$ for $i$, given its type $\theta_i$, is:

$$u_i(\sigma_i|\theta_i, \sigma_{-i}) = \sum_{\theta_{-i} \in \Theta_{-i}} \Pr(\theta_{-i}|\theta_i) u_i(\langle \sigma_i(\theta_i), \sigma_{-i}(\theta_{-i}) \rangle, \theta_i) \quad (1)$$

where the utility of a mixed strategy profile is defined in the obvious way. A *best response* for $i$ to a strategy profile $\sigma_{-i}$ is any $\sigma_i$ that maximizes Eq. (1) for each of its types $\theta_i$. A strategy profile $\sigma \in \Sigma$ is a *Bayes-Nash equilibrium (BNE)* iff $\sigma_i$ is a best response to $\sigma_{-i}$ for each $i$. Intuitively, a BNE is stable in the sense that each agent is maximizing its expected utility given the strategies of all other agents, where the expectation is taken with respect to possible realizations of the other agents's types. A *dominant strategy* for agent $i$ is a strategy $\sigma_i$ that maximizes Eq. (1) no matter what strategies are played by the others. A *dominant strategy equilibrium (DSE)* is a strategy profile consisting of dominant strategies for each agent $i$.

### 2.2 Mechanism Design

Mechanism design deals with the problem of designing a game—in which a collection of self-interested agents interact—so as to optimize some objective on the part of the designer [13]. More formally, we have a collection of agents (e.g., potential buyers of some good) and some set of outcomes $O$ (e.g., the allocation of a good to a particular agent at a specific price). As above, each agent $i$ has a type $\theta_i \in \Theta_i$ known only to itself and utility function $u_i$, where $u_i(o, \theta_i)$ reflects the utility of outcome $o$ to agent $i$ if its type is $\theta_i$. For example, the type of the agent might dictate its valuation for the good being auctioned, with its utility for any outcome $o$ in which it obtains the good given by its valuation less the price paid.

The mechanism designer has some *social choice function* $f$ it wishes to optimize (say, maximize), where $f(o, \theta)$ denotes the objective value (to the designer) of outcome $o$ when the agent type profile is $\theta$. For example, the designer may wish to maximize social welfare by ensuring the good goes to the agent with the highest valuation. A *deterministic mechanism* comprises a set of actions $A_i$ for each agent $i$, and an outcome rule $g : A \to O$ mapping action profiles $A$ into outcomes. A *randomized mechanism* associates each action profile with a distribution over outcomes. As such, a mechanism induces a Bayesian game among the agents. A mechanism *implements* a social choice function $f$ iff, in equilibrium, the outcome of the game is $o \in \arg\max f(o, \theta)$ whenever the agents types are given by $\theta$. The mechanism design problem is to find a mechanism that implements $f$ using the desired notion of equilibrium (most commonly, BNE of DSE), possibly subject to various constraints.

The *revelation principle* makes the mechanism design problem somewhat simpler by noting that if a mechanism exists that implements $f$, then a direct, incentive-compatible mechanism exists for $f$ as well. In other words, we let strategies correspond to types (hence agents directly reveal their types), and in equilibrium, each agent will *truthfully* reveal its type. The revelation principle has led to an almost exclusive focus on direct, incentive-compatible mechanisms.

As mentioned above, the use of direct mechanisms in realistic domains faces the problem that revelation of utility functions is quite often impractical. For example, in large, multi-attribute outcome spaces, simply communicating a utility function may be problematic unless it has considerable structure. Recent work on preference elicitation addresses this issue in the context of combinatorial auctions [7]. Even when utility functions are compactly representable, computing precise values, or eliciting these from humans, may prove problematic. The problem of mechanism design under such circumstances has recently been addressed in the context of auctions. Blumrosen and Nisan [1] propose a model for limited-precision bids in auctions that is designed to deal with communication complexity, and devise dominant strategy mechanisms in which agents (implicitly) reveal upper and lower bounds on their valuations. Parkes [18] addresses the problem of agents with uncertain valuations facing the decision of whether to pay the computational cost of refining their utility estimates when participating in an auction. Larson and Sandholm [12] study similar phenomenon in bargaining settings.

Much work in mechanism design deals with mechanisms for very specific settings. However, some general families of mechanisms have been constructed that apply rather widely (e.g., the celebrated Vickery-Clarkes-Grove mechanism) assuming certain restrictive, though reasonable conditions and social choice functions (e.g., the possibility of payments, quasi-linear utility, and social welfare maximization) without requiring assumptions about the specific priors. Conversely, direct optimization of the objective (subject to certain constraints) in a way that exploits the prior can lead to a more algorithmic mathematical programming framework for mechanism design. Myerson's [15] approach to revenue-optimal auction design is the standard exemplar, giving rise to a class of mechanisms that can be tailored algorithmically to any specific prior. In *automated mechanism design* [8, 20] the specific details of the problem—outcome space, social choice function, and, critically, the *prior*—are taken as input, and one automatically constructs a mechanism by formulating an optimization problem that maximizes the social choice function subject to certain constraints. These constraints specify the equilibrium concept to be used (BNE, DSE), various rationality constraints, etc. Conitzer and Sandholm [8] show how randomized mechanisms in finite settings (specifically, those with finite type spaces) can be optimized using linear programming.

### 2.3 Minimax Regret

The problems of utility elicitation facing direct mechanisms obviously arise in non-strategic decision making contexts as well. Research on preference elicitation for single-agent decision making must often address the problem of making decisions with incompletely or imprecisely specified utility functions. While *minimax regret* is a common criterion for decision making under strict uncertainty [21, 10, 2], only recently has it been proposed as a means for dealing with imprecisely specified utility [3, 19]. We briefly overview the notion as applied to imprecise utility.

We assume a set of possible decisions $D$ from which a specific decision must be taken on behalf of, or recommended to, some user (say, by an automated decision system). A utility function $u$ associates an expected utility $u(d)$ with each $d \in D$. However, the system is unsure of the user's utility function, knowing only that it lies in some feasible set $U$. For example, the system may have elicited bounds on the user's utility for various outcomes, and $U$ represents the set of utility functions consistent with the bounds elicited thus far. We define the *regret* of $d$ w.r.t. $u$ to be

$$R(d, u) = \max_{d' \in D} u(d') - u(d).$$

This reflects the loss experienced by taking decision $d$ instead of acting optimally when the true utility function is

$u$. The *max regret* of $d$ w.r.t. $U$ is

$$MR(d, U) = \max_{u \in U} R(d, u),$$

reflecting the worst-case regret of $d$ should an adversary be allowed to choose the user's utility function from feasible set $U$. Finally, the *minimax optimal decision* w.r.t. $U$ is that $d$ with minimal max regret:

$$d_U^* = \arg\min_{d \in D} MR(d, U).$$

While conceptually straightforward, direct computation of minimax regret is often not feasible, because of the complexity of the decision and utility spaces. As a consequence, when applied to practical problems, care must be taken to exploit computationally whatever structure (e.g., conditional utility independence, graphical action models, etc.) exists in the problem (see [4, 5, 6, 24] for further motivation and discussion of computational issues).

## 3 Games with Strict Type Uncertainty

In this section we define incomplete information games with strictly uncertain priors. We propose the use of minimax regret as a decision criterion for participants in such a game, define minimax-regret equilibria, and prove that minimax-regret equilibria exist in mixed strategies.

### 3.1 Definition

An *incomplete information game with strict type uncertainty* consists of the same components as a (Bayesian) incomplete information game, but has a *qualitative* or *strict prior* rather than a probabilistic prior. Specifically, we assume an action set $A_i$, type space $\Theta_i$, and utility function $u_i$ for each agent $i$. We assume a *strict prior* $T \subseteq \Theta$ representing (common) beliefs about possible type profiles held by the agents in the game. Intuitively, $\Theta$ denotes the set of possible types from a "structural" perspective, while $T$ denotes what is believed: only type profiles $\theta \in T$ are considered to be possible given the information possessed by the participants. While we could simply treat $\Theta$ itself as the set of credible types, when we discuss partial and incremental revelation mechanisms, this distinction will be useful. Strict incomplete information games are equivalent to games in *informational form* [11].

As in the standard setting, we assume each agent knows its own type. Its beliefs about the types of other agents, given its type $\theta_i$, is given by the set

$$T(\theta_i) = \{\theta_{-i} : \langle \theta_i, \theta_{-i} \rangle \in T\}$$

(i.e., those type profiles consistent with its own known type). Strategies are defined in the usual way as mappings from types to (mixed) action choices.

### 3.2 Minimax-regret Equilibrium

The expected utility of a fixed strategy $\sigma_i$ as defined in Eq. (1) requires some distribution over the possible types of other agents. Without distributional information, we must adopt some qualitative decision criterion to evaluate and compare strategies. Here we propose the use of the minimax-regret decision criterion.

**Definition 1** The *regret* of strategy $\sigma_i$ for agent $i$ with type $\theta_i$, given strategy profile $\sigma_{-i}$ and type profile $\theta_{-i}$ of the other agents, is

$$\begin{aligned} R_i(\sigma_i | \theta_i, \theta_{-i}, \sigma_{-i}) = \\ \max_{\sigma_i' \in \Sigma_i} [u_i(\langle \sigma_i'(\theta_i), \sigma_{-i}(\theta_{-i}) \rangle, \theta_i) \\ - u_i(\langle \sigma_i(\theta_i), \sigma_{-i}(\theta_{-i}) \rangle, \theta_i)]. \end{aligned} \quad (2)$$

The *max regret* of strategy $\sigma_i$ w.r.t. prior $T$, given $\theta_i$ and $\sigma_{-i}$ is

$$MR_i(\sigma_i | \theta_i, T, \sigma_{-i}) = \max_{\theta_{-i} \in T(\theta_i)} R_i(\sigma_i | \theta_i, \theta_{-i}, \sigma_{-i}). \quad (3)$$

Finally, a *minimax best response* of agent $i$ to $\sigma_{-i}$ w.r.t. $T$ is any strategy $\sigma_i^*$ satisfying, for all $\theta_i \in \Theta_i$:

$$\sigma_i^* \in \arg\min_{\sigma_i \in \Sigma_i} MR_i(\sigma_i | \theta_i, T, \sigma_{-i}). \quad (4)$$

Intuitively, if we fix the behavior and types of all other agents, the regret of agent $i$ with type $\theta_i$ for playing $\sigma_i$ is the loss $i$ experiences by playing $\sigma_i$ rather than acting optimally. Of course, agent $i$ does not know the true types of the other agents. The max regret of $\sigma_i$ given prior $T$ is the most $i$ could regret playing $\sigma_i$ (against the fixed strategies of the others) should an adversary choose its opponents's types in a manner consistent with its beliefs. Finally, a minimax best response is any strategy that minimizes this worst case loss in the face of such an adversary. Note that this strategy requires a minimax optimal choice for every possible type agent $i$ could possess.

Unlike standard best responses, minimax best responses require agents to adopt a cautious stance with respect to possible realizations of opponent types. Without probabilistic information quantifying type uncertainty, minimax regret seems like the most natural decision criterion that could be adopted by such agents.

We define the notion of a *minimax-regret equilibrium* for a strict incomplete information game by analogy with Bayes-Nash equilibrium.

**Definition 2** A strategy profile $\sigma$ is a *minimax-regret equilibrium* iff $\sigma_i$ is a minimax best response to $\sigma_{-i}$ for all agents $i$.

We note that other notions of qualitative equilibria have been proposed, but none have the same flavor as minimax-regret equilibria. Tennenholtz [23] describes qualitative equilibria for complete information games that rely on maximin strategies; but these do not have a clear extension to incomplete information games with type uncertainty. Work on *uncertainty aversion* can be viewed as incorporating some form of strict type uncertainty, but in a very different way. Rather than truly qualitative uncertainty, each agent is assumed to have a *set* of probabilistic priors (thus combining qualitative and quantitative uncertainty) [22]. Recently, equilibrium analysis of various auctions has been considered using this notion [9, 16]. Analysis of games in informational form naturally bears the closest relation to our work; however, to date, only *ex post* equilibria have been proposed for such games [11], which are considerably stronger than minimax-regret equilibria, and are not guaranteed to exist.

Dominant strategies for strict incomplete information games can be defined in a similar way: we say $\sigma_i$ is *minimax-dominant* if it is a minimax best response for *any* strategies $\sigma_{-i}$ adopted by other players.[1] A *minimax-dominant strategy equilibrium (minimax-DSE)* is any strategy profile consisting of minimax-dominant strategies.

### 3.3 Existence of Equilibria

Not surprisingly, pure strategy minimax-regret equilibria for strict incomplete information games do not always exist (as is the case for BNE in Bayesian games). However, we can show that mixed-strategy minimax-regret equilibria exist for any finite game.

**Theorem 1** *A mixed strategy minimax-regret equilibrium exists for any strict incomplete information game with finite agent, action, and type spaces.*

*Proof sketch:* The result can be proved using a similar strategy to classic proofs of the existence of (Bayes) Nash equilibria for finite games. We use Kakutani's fixed point theorem to show that the minimax-best-response correspondence (i.e., the mapping from any strategy profile to the set of profiles obtained by composing individual best responses to it) has a fixed point—this, by definition, is a minimax equilibrium. To apply the theorem, we show that minimax-best-response set for any profile is convex, and that the correspondence is upper-hemicontinuous. This relies on the piecewise-linear, convex nature of the max regret function itself (thus having a rather different character than best response correspondences based on expected utility). We provide full details in a longer version of the paper.

The characterization of conditions under which minimax-regret equilibria exist when type spaces are continuous is of obvious interest. We expect that similar characterizations for Bayesian games might be applicable with suitable modifications [14].

Because dominant strategies in Bayesian games do not rely on the precise form on the prior, but only the set of possible types, we have (not surprisingly):

**Proposition 1** *Strategy profile $\sigma$ is a minimax-DSE for a strict incomplete information game with prior $T \subseteq \Theta$ iff it is a DSE for Bayesian game with type set $T$.*

## 4 Minimax-based Mechanisms

We now consider the setting in which a mechanism designer is faced with a mechanism design problem in which prior information over types available to the designer and the participants cannot be characterized probabilistically. In other words, type uncertainty is strict. There are many cases in which probabilistic priors over types may be difficult or inconvenient to assess, whereas qualitative information may be relatively easy to obtain. For example, when requesting valuation information in an auction, it may be much more natural to maintain upper and lower bounds than distributional information. Since the mechanism designer is faced with the same strict uncertainty as the participants, the mechanism cannot be realized by optimizing the social choice function using the *expected value* induced by the mechanism with respect to realization of participant types. We instead propose to view the mechanism designer as a regret minimizer as well.

If we restrict our attention in the typical fashion to direct mechanisms, then—even with strict type uncertainty—some general *prior independent* mechanisms such as VCG can be applied when the setting allows (e.g., if we allow payments and are interested in maximizing social welfare). But notice that approaches that optimize with respect to specific (families of) probabilistic priors [15, 8] cannot be applied in the case of strict type uncertainty. Even though the mechanism may be able to induce participants to reveal their types $\theta$ truthfully, it may not be able to simply maximize $f(o, \theta)$ without violating constraints such as incentive compatability. Thus the designer may regret the choice of one incentive compatible mechanism relative to another depending on the specific realization of agent types. Rather than maximizing the expected value of $f$ subject to the typical constraints, we approach the problem by minimizing the max regret of the mechanism.

The importance of minimax-regret equilibria can be appreciated more fully when considering mechanism design settings in which full types cannot be practically revealed. In such a case, the mechanism is forced to optimize $f$ without full type information; without a probabilistic prior it cannot use expectations over types given the partial information obtained. To emphasize the added importance of regret minimization on the part of the mechanism, in this section we describe strict mechanism design problems and how a mechanism might handle *partial revelation* of types.

---

[1] One could alternatively define dominant strategies in a "prior independent" way by requiring that regret be defined w.r.t. *any* type vector in $\Theta_{-i}$. This would be somewhat more consistent with the typical definition in Bayesian games. This version will prove useful later.

### 4.1 Strict Mechanism Design Problems

A *strict mechanism design* problem is a mechanism design problem—comprising agents $i \leq N$, outcomes $O$, types $\Theta_i$, utilities $u_i$ and social choice function $f$—in which the prior $T \subseteq \Theta$ reflects strict uncertainty. A mechanism $M = \langle A, g \rangle$ is defined in the standard way, as a set actions $A_i$ for each player and outcome function $g : A \mapsto O$ if the mechanism is deterministic, or $g : A \mapsto \Delta(O)$ if it is randomized. Note that a mechanism $M$ induces a strict incomplete information game.

We say $M$ implements $f$ if, given any type profile $\theta$, the only action profiles $a$ taken (with positive probability) in equilibrium are such that $g(a) \in \arg\max_{o \in O} f(o, \theta)$, possibly subject to certain constraints.[2] Since $M$ induces a strict incomplete information game, we can't use BNE as our implementation solution concept; but both minimax-regret equilibria and DSE can be adopted.

When restricting attention to direct mechanisms, the action space is simply $A_i = \theta_i$: each agent reports its type (possibly untruthfully) to the mechanism. Incentive compatibility of a strict mechanism can be defined in the standard way. More interesting is the case where we don't rely on full type revelation. We define a *direct, partial revelation (DPR)* mechanism as follows: we assume an action set $S_i$ for each agent $i$, with each $s_i \in S_i$ a subset $s_i \subseteq \Theta_i$ of possible types. Intuitively, $i$'s taking action $s_i$ is interpreted as a "claim" that $\theta_i \in s_i$.[3] Note that direct mechanisms are a special case of DPR mechanisms. We will generally assume that $\cup S_i = \Theta_i$. When this is the case, we can define truthful strategies: agent $i$ only takes action $s_i$ when $\theta_i \in s_i$. An *incentive-compatible mechanism* in the DPR case is one in which all agents act truthfully in equilibrium.

In rich settings with a sufficiently sparse action space (i.e., severe restrictions on the amount of revelation), it will not generally be possible to maximize $f$ (subject to relevant constraints). Instead of insisting on a faithful implementation of $f$ (i.e., ensuring that we always obtain an outcome $o \in \arg\max f(o, \theta)$ whenever the type vector is $\theta$), we might instead require that we maximize the *expected value* of $f$ relative to $\theta$ given whatever intuitive constraints we wish to place on the mechanism (e.g., incentive compatibility, limited revelation, etc.). This is similar in spirit to "prior-specific" mathematical programming approaches [15, 8], where the goal is to produce a mechanism that maximizes $f$ subject to certain constraints.[4]

Once again, though, strict type uncertainty along with partial revelation prevent the mechanism designer from maximizing the *expected* value of $f$. Instead, we can consider the regret of the mechanism designer: the designer wishes to find a mechanism that, in equilibrium, minimizes regret over $f$ with respect to possible realizations of the agents's types. Note that we now consider both the regret of the agents participating in the mechanism—agents consider the regret of their actions w.r.t. their own expected utility and adopt a minimax-regret equilibrium (or DSE) given the rules of the mechanism—and of the designer—it adopts a mechanism that minimizes regret w.r.t. the value of the social choice function, assuming the agents play a minimax-regret equilibrium (or DSE).

### 4.2 Automated Mechanism Design

In the spirit of AMD, we provide a formulation of the optimization problem for a finite-type, finite-outcome, strict mechanism design problem in which the goal is to find an incentive compatible DPR mechanism satisfying ex-post rationality.[5] Again, we emphasize that direct mechanisms are a special case of DPR mechanisms; and these may still require regret minimization due to constraints imposed on the mechanism.

Let $T \subseteq \Theta$ be a common prior over type profiles. Let $S_i$ be the set of possible "partial types" (or subsets) $i$ can reveal, with $S = \times_i S_i$. We'll assume for ease of presentation that the distinct reports for each $i$ are mutually exclusive (i.e., $s_i \cap s'_i = \emptyset$ for any two different reports). Given this fixed action space, any mechanism can be specified by a collection of parameters $\langle p^o_s : o \in O, s \in S \rangle$, where $p^o_s$ denotes the probability of outcome $o$ being realized when partial-type profile $s$ is revealed. Let $T(s)$ be the collection of type profiles in $T$ consistent with the partial-type profile $s$; define $\Theta(s)$ similarly. Conversely, let $S(T)$ be the set of partial-type profiles that correspond to some element of $T$ (similarly for $S(\Theta)$), and let $s(\theta)$ be the unique partial-type profile consistent with $\theta$. The notation $T_i(s_i)$, $S_i(\Theta_i)$, $s_i(\theta_i)$, etc. is defined in an analogous fashion.

#### 4.2.1 Formulation of Mechanism Optimization

Restricting attention to incentive compatible mechanisms (a restriction that must be enforced via constraints on the optimization), we define the pairwise regret of mechanism $\mathbf{p} = \langle p^o_s \rangle$ w.r.t. mechanism $\mathbf{q} = \langle q^o_s \rangle$ to be

$$MR(\mathbf{p}, \mathbf{q}) = \max_{\theta \in T} \sum_o (q^o_{s(\theta)} - p^o_{s(\theta)}) f(o, \theta) \qquad (5)$$

Intuitively, the regret of mechanism $\mathbf{p}$ relative to $\mathbf{q}$ is the maximal loss in $f$ incurred by adopting $\mathbf{p}$, allowing an ad-versary to choose the agents's types $\theta$ and their (truthful) reports $s$.[6] Mechanism **p** is *minimax optimal* if it satisfies

$$\mathbf{p} \in \arg\min_{\mathbf{p}} \max_{\mathbf{q}} MR(\mathbf{p}, \mathbf{q}). \qquad (6)$$

---

[2] If $g$ is randomized, then we require that $g(a, o) > 0$ only if $o$ is a maximizing outcome.

[3] Thus, $S_i$ can be viewed (indirectly) as a *query language* [18], though we do not consider incremental querying in this paper.

[4] None of these approaches consider partial revelation. As discussed above, the Myerson auction is prior-specific only in the sense that the general mechanism "template" is instantiated to produce a different concrete mechanism by plugging in any specific prior.

[5] With the possibility of payments, the outcome space is no longer finite; but this can be dealt with separately (see below).

The optimization in Eq. (6) can be formulated in different ways by imposing different constraints on the mechanisms in question. For instance, we might insist that our chosen mechanisms **p** and **q** be incentive compatible and satisfy *ex post rationality*; we formulate this next. But other restrictions are possible, and one could even formulate regret using different restrictions on **p** and **q** (e.g., by allowing the adversary to consider a wider space of mechanisms **q** than the designer can for **p**, or even allowing it to make "perfect" outcome choices, thus giving it more power still).

Converting the minimax program that encodes Eq. (6) into a minimization leads to the following formulation:

$$\min_{\delta, \mathbf{p}} \delta$$

$$\text{s.t.} \sum_o (q_s^o - p_s^o) f(o, \theta) \leq \delta \quad \forall \mathbf{q}, s, \theta \in \Theta(s) \qquad (7)$$

$$R_i(s_i, s_i'|\theta_i, \theta_{-i}, \mathbf{p}) \leq \min_{d \in \Delta(S_i)} MR_i(d|\theta_i, T_{-i}, \mathbf{p})$$

$$\forall i, s_i, s_i', \theta_i \in T_i(s_i), \theta_{-i} \in T_{-i} \qquad (8)$$

$$\sum_o p_s^o u_i(o, \theta_i) \geq 0 \quad \forall i, s_i, \theta_i \in T_i(s_i) \qquad (9)$$

$$\sum_o p_s^o = 1 \quad \forall s \qquad (10)$$

$$p_s^o \geq 0 \quad \forall s, o \qquad (11)$$

The variables in this program are parameters $p_s^o$ of the mechanism **p** being optimized (constrained by the obvious simplex constraints (10) and (11)), and $\delta$, measuring the minimax regret of **p**. The aim then is to minimize $\delta$ subject to the constraints above. We now describe the role of each of the constraint sets (7–9) in turn.

The (infinite) collection of *regret constraints* (7) ensures that the regret of **p**, *from the mechanism designer's perspective*, with respect to any alternative mechanism **q** is less than $\delta$. This therefore "defines" $\delta$ as the max regret of **p**. This effectively replaces the maximization over **q** in Eq. (6) by universal quantification. We will see below how constraints on the space of "adverarial" mechanisms **q** are taken into account.

The (nonlinear) *IC-constraints* (8) ensure incentive compatibility of the mechanism **p**. Here we define

$$R_i(s_i, s_i'|\theta_i, \theta_{-i}, \mathbf{p}) =$$

$$\sum_o (p_{s_i', s_{-i}(\theta_{-i})}^o - p_{s_i, s_{-i}(\theta_{-i})}^o) u_i(o, \theta_i)$$

---

[6]If we allow "overlapping" reports in which several reports correspond to the same type, we can extend this definition by (adversarially) maximizing over $s \in S(\theta)$.

which is the pairwise regret of agent $i$ should it report $s_i$ instead of $s_i'$ in mechanism **p** when its true type is $\theta_i$ and others have type $\theta_{-i}$ (and truthfully report $s_{-i}(\theta_{-i})$). The right-hand side of the constraint refers to alternative (randomized) reporting strategies $d \in \Delta(S_i)$ that $i$ could adopt, and their maximum regret in mechanism **p** (we address this below). By quantifying over all types $\theta_i$ that "truthfully" correspond to the reported partial type $s_i$, we ensure that agents have incentive to report truthfully (i.e., the max regret of reporting truthful partial type $s_i$ w.r.t. **p** is no greater than the max regret of any randomized strategy). Notice that this defines the value of truthful reporting in a minimax-regret equilibrium. It is possible to insist on a minimax-DSE in an entirely analogous fashion.

The *EPR-constraints* (9) enforce *ex post rationality*: each $i$ will participate in **p** even knowing the true types of the other players, since its expected utility for participating is non-negative (we assume a baseline utility of zero for non-participation). These IC and EPR constraint sets are analogous to those proposed in [8], with expected utility replaced by regret minimization.

Given this formulation, there are two key practical difficulties in solving this optimization problem. First, the set of regret constraints (7) is infinite, preventing us from direct optimization as a finite mathematical program. Second, the IC-constraints (8) are nonlinear due to the presence of the minimization on the RHS of each of the constraints. Note however that all other constraints are linear. We address each of these two problems in turn, resulting in an algorithm for regret-based mechanism design involving only the solution of linear, mixed-integer programs.

### 4.2.2 Constraint Generation

Practically, we deal with the infinite number of regret constraints by constraint generation. Specifically, we solve a relaxed version of the program by considering only a finite subset of the constraints and linearizing any IC constraints (see below). Given a relaxed solution $\langle \mathbf{p}, \delta \rangle$, we then find the maximally violated regret constraint by finding the mechanism parameters **q** and partial type report $s$ and type profile $\theta$ that maximize the regret of **p**. This is a mixed integer quadratic program:

$$\max_{(I_\theta), \mathbf{q}} \sum_o (q_{s(\theta)}^o - p_{s(\theta)}^o) I_\theta f(o, \theta)$$

$$\text{s.t. IC, EPR and simplex constraints on } \mathbf{q} \qquad (12)$$

$$\sum_\theta I_\theta = 1 \qquad (13)$$

where the $I_\theta$ are boolean variables indicating which type profile $\theta$ (and therefore which $s$) is chosen. Here the constraints are either linear or can be linearized (see below) but the objective is quadratic. This optimization can be reformulated as a mixed integer *linear* program (MIP) as

follows:

$$\max_{(I_\theta), \mathbf{q}, (Y_\theta^o)} \sum_o Y_\theta^o f(o, \theta)$$

s.t. IC, EPR and simplex constraints

$$I_\theta \cdot lb \leq Y_\theta^o \leq I_\theta \cdot ub \quad \forall o, \theta \quad (14)$$

$$Y_\theta^o \leq (q_{s(\theta)}^o - p_{s(\theta)}^o) - lb \cdot (1 - I_\theta)$$
$$\forall o, \theta \text{ where } f(o, \theta) \geq 0 \quad (15)$$

$$Y_\theta^o \geq ub \cdot (1 - I_\theta) - (q_{s(\theta)}^o - p_{s(\theta)}^o)$$
$$\forall o, \theta \text{ where } f(o, \theta) < 0 \quad (16)$$

where $lb$ and $ub$ are bounds on the value of $(q_{s(\theta)}^o - p_{s(\theta)}^o)$ (i.e., $-1$ and $1$). The trick here is to introduce new variables $(Y_\theta^o)$ to replace the product in the quadratic objective and to add constraints that force those variables to 0 if they don't correspond to the chosen $\theta$ (14) or to the true value of the original product if they do (15 and 16).

If the regret induced by $\mathbf{q}$ exceeds $\delta$, then this regret constraint is violated at the proposed solution, so we add the constraint (7) corresponding to $\mathbf{q}$, $s$, and $\theta$. The optimization solved to generate mechanism $\mathbf{q}$ can involve the same restrictions (e.g., IC, EPR) as those for $\mathbf{p}$, or we may relax these to give the adversary more power.

### 4.2.3 Constraint Linearization

We adopt a similar strategy for dealing with the nonlinear IC constraints. The min on the right-hand side of constraint (8) cannot be expressed linearly. So given the current solution $\mathbf{p}$, we compute the stochastic report $d$ for each agent $i$, and each feasible type $\theta_i$, that has minimax regret in mechanism $\mathbf{p}$ given that $i$ has type $\theta_i$. This can be solved using a straightforward linear minimization if we enumerate all types (or we can use constraint generation to prevent doing so). To determine whether any of the IC constraints for $i, \theta_i$ are violated, we can either explicitly compute $R_i(s_i, s_i'|\theta_i, \theta_{-i}, \mathbf{p})$ for $s_i = s_i(\theta_i)$, and all $s_i'$ and $\theta_{-i}$, or we can simply maximize $R_i$ (as an LP). Finally, we can test whether any of these constraints is violated by comparing the maximum $R_i$ value obtained with the value of the solution of the min on the right-hand side of the constraint. If a violation exists, we add the maximally violated constraint (or all constraints corresponding to the agent and type in question) to tighten the master LP.

Once the maximally violated (regret and IC) constraints have been added to the LP, we re-solve the tightened LP to obtain a new solution $\mathbf{p}$. When we get to a point where no violated constraints are found, we are assured that the minimax optimal mechanism has been computed. This approach reduces the entire optimization problem to a sequence of LPs and MIPs.

### 4.2.4 Payments

If one allows payments among participants (one could hardly design an auction without payments), the transfer of money can be encoded within the outcome space $O$. However, this will render the outcome space infinite and make the finite optimization above impossible (at least in this explicit form). However, under the assumption of quasi-linear utility, we can treat payments separately from the outcome space itself, as is standard in mechanism design. The method above can then easily be adapted by including variables for payments that are separate from the outcome variables. In such settings, optimization criteria such as revenue maximization can be considered, as can constraints such as budget balance or no deficit.[7]

### 4.3 Experimental Results

We have experimented with this approach to automated strict mechanism design in the special case of direct mechanisms (i.e., complete revelation) with IC and EPR. We consider two objective functions—social welfare and revenue (when payments are permitted)—and two types of adversaries—an "omniscient" adversary that can pick the optimal outcome for each revealed type vector without regard to whether the players would actually reveal truthfully, or a "constrained" adversary that picks, for each possible type vector, an IC, EPR mechanism that maximizes the objective function. In the case of mechanisms with payments, we have experimented with various payment constraints: budget balance (sum is zero), no deficit (sum is non-negative) or unconstrained.

The first problem we consider is a strictly uncertain variant of the simple two-player, two-type "Divorce Settlement" problem [20], in which an arbitrator decides among one of four options for dealing with a jointly owned painting in a divorce: the husband gets the painting; the wife gets it; it is hung in a museum; or it is burned. The husband and wife each have two possible types, low (less attached to the painting) and high (more attached). The utility function of the low type is: 2 (get it), 0 (other gets it); 1 (museum); -10 (burn). For the high type, utility is 100, 0, 50 -10.

Figure 1 shows the max-regret minimizing mechanism under social welfare (with no payments). This mechanism is identical for "omniscient" and "constrained" adversaries.[8] The max regret of this mechanism is 27.7. Note that this mechanism, unlike those derived in [20], is symmetric, treating the husband and wife equally.

When maximizing social welfare with payments (regardless of payment constraints) a zero-regret mechanism can be derived (in the no deficit case this is not surprising since a VCG mechanism would optimize welfare). The mechanism always gives the painting to the husband, except when the wife's utility for it is strictly higher; and if a party with high type receives the painting, it must pay the designer just enough to remove any incentive to lie. When maximizing revenue, the adversary is always able to extract payments equal to the highest utility of the agents. The mechanism designer however cannot do as well (with an IC-EPR mechanism) as the adversary, but it is interesting to note that being allowed to run a deficit actually helps reduce regret. Figures 2 and 3 show the mechanisms for both the No Deficit and Deficit Allowed cases. In the former, the designer can extract a payment of $55.4545$ from the party receiving the painting, except when both have low types, in which case no payment is required. In the latter case, taking the risk of giving money to both agents in the Low/Low profile enables the designer to extract a higher payment when at least one has high type. The regret of $44.546$ in the No Deficit case is significantly reduced to $15.183$ by allowing a deficit.

---

[7]Our formulation is generalized as in AMD [8].

[8]In all problems described here, these two cases lead to the same result: for each type vector, there is an IC-EPR mechanism with the same objective value as the optimal outcome choice of the omniscient adversary.

|   | Low | High |
|---|---|---|
| L | 0.5/0.5<br>0/0 | .2273/.7273<br>0/.0455 |
| H | .7273/.2273<br>0/.0455 | .4545/.4545<br>0/.0909 |

Figure 1: Regret-minimizing mechanism for the Divorce Settlement problem (social welfare, no payments). Each cell shows the probability of each of the four outcomes as a function of the report of the husband (row) and wife (column): husband/wife at the top and museum/burn at the bottom of each cell.

|   | Low | High |
|---|---|---|
| L | .4545/.4545<br>0/.0909 | 0/1<br>0/0 |
| H | 1/0<br>0/0 | 0/1<br>0/0 |

|   | Low | High |
|---|---|---|
| L | 0/0 | 0/55.4545 |
| H | 55.4545/0 | 0/55.4545 |

Figure 2: Divorce Settlement (max revenue, payments, no deficit). Top matrix shows outcome probabilities, the bottom matrix payments.

We have also experimented with our approach with one-item, two-agent auctions, with a finite set of possible valuations. We experimented with five types per agent (with valuations $0$, $0.25$, $0.5$, $0.75$ and $1$) following the example of [20]; and also with three types (with valuations of $0.25$, $0.5$ or $0.75$), a subset of the original five types. The minimax regret levels are summarized in Table 1 for mechanisms with payments and a No Deficit constraint, against a "constrained" adversary. Naturally, the regret of the three-type mechanisms are lower since there is "less uncertainty". Note that regret reduction strongly suggests how incremental type elicitation could be used in the design of mechanisms. When optimizing for social welfare, a zero-regret mechanism is achieved (not surprisingly, since a second-price auction can maximize social welfare with dominant strategies).

|   | Low | High |
|---|---|---|
| L | .1549/.1549<br>0/.6901 | 0/1<br>0/0 |
| H | 1/0<br>0/0 | 0/1<br>0/0 |

|   | Low | High |
|---|---|---|
| L | $-6.5915/-6.5915$ | 0/84.8169 |
| H | 84.8169/0 | 0/84.8169 |

Figure 3: Divorce Settlement (max revenue, payments, deficit allowed). Top matrix shows outcome probabilities, the bottom matrix payments.

|   | Social Welfare | Revenue |
|---|---|---|
| Auction-5 | 0 | 0.3264 |
| Auction-3 | 0 | 0.1797 |

Table 1: Regret for the five- and three-type auctions using social welfare and revenue maximization (under No Deficit constraint).

A number of approaches to scaling the automated optimization of mechanisms in settings of strict uncertainty could be adopted. The optimization needed is more complex than in standard AMD because of the minimax optimization criterion. Further empirical study is needed to gauge the complexity and convergence properties of the constraint generation process. Apart from that, however, the scaling issues facing strict AMD are similar to those facing AMD in the expected utility context; and many of the solutions will be similar. We discuss some of these issues in the next section.

## 5 Concluding Remarks

We have introduced strict incomplete information games as a variant of Bayesian games in which type uncertainty is strict. We proposed minimizing max regret with respect to possible realizations of other agent types as a natural optimization criterion for players in such games, and defined minimax equilibria. We also argued that strict uncertainty is natural in many mechanism design settings; and that mechanisms themselves, especially when forced to rely on partial type revelation, must deal with strict type uncertainty when deciding on outcomes. We showed how the minimax regret optimization of mechanisms can be reduced to a series of LPs and MIPs in certain circumstances.

This work forms a starting point for further investigation into mechanism design with incremental or partial type revelation. As argued earlier, in these settings semi-qualitative decision criteria such as minimax regret will often be most appropriate from the perspective of mechanism optimization. A number of interesting directions remain to be pursued. We are interested in developing more general mechanisms (like VCG) involving partial or incremental elicitation under strict type uncertainty. The specific proposals for strict AMD suggested here need to be further explored empirically. Along these lines, we hope to explore the approximate optimization of mechanisms.

Of interest also are issues that face (and are being addressed in) classical AMD, for example, exploiting structure in the action space, outcome space, and type spaces of agents for computational benefit. Dealing with continuous type spaces is critical for practical problems, since utility functions are usually continuously parameterized. Notice that finite message mechanisms can deal with continuous type spaces effectively through partial revelation (e.g., using the kinds of partial-type mappings suggested here). It should be relatively straightforward to extend our approach to minimax mechanism optimization to such settings. Within this context, we might also consider methods for optimizing the "meaning" of reported partial types, sequential revelation of types, and the means to address the tradeoff between communication cost and mechanism value.

**Acknowledgments**

Thanks to Moshe Tennenholtz for very helpful suggestions, David Parkes and Tuomas Sandholm for discussions, Rakesh Vohra for pointers to relevant literature, and Pascal Poupart for suggesting the objective linearization in Sec. 4.2.2. This research was supported by the Institute for Robotics and Intelligent Systems (IRIS) and the Natural Sciences and Engineering Research Council (NSERC).